\documentclass[
    ,final            % use final for the camera ready runs
  ]
  {aipproc}

\usepackage{color}
\layoutstyle{8x11double}

\newcommand{\Msun}      {\mbox{$\rm\,M_{\mathord\odot}$}}

\begin{document}

\title{The Evolution of Accreting Black Holes in Outburst}

\author{John A. Tomsick}{
  address={Center for Astrophysics and Space Sciences, Code 0424, 
University of California at San Diego, La Jolla, CA 92093; jtomsick@ucsd.edu}}

\begin{abstract}

Black hole binaries exhibit dramatic changes in their X-ray 
spectral and timing properties over time, providing important clues about 
the physical processes that occur in these systems.  Black holes and black 
hole candidates are prime targets for {\em RXTE} with observational goals 
including the study of extreme gravitational fields and jet formation 
mechanisms.  The great wealth of data from {\em RXTE} has helped us to 
learn about these systems as well as raising new questions about accreting 
black holes.  {\em RXTE} observations have allowed us to study a wide range 
of black hole science topics including the connection between the accretion 
disk and jets, the geometry of the inner accretion flow, and the physical 
changes that occur between spectral states.  In this presentation, I discuss 
significant results on these topics that have been obtained for persistent 
and transient black holes over the past several years, and I present results 
from our program of X-ray and radio observations during the decays of black 
hole transient outbursts.

\end{abstract}

\maketitle

%%%%%%%%%%%%%%%%%%%%%%%%%%%%%%%%%%%%%%%%%%%%
%% MAINMATTER
%%%%%%%%%%%%%%%%%%%%%%%%%%%%%%%%%%%%%%%%%%%%

\section{Objective}

The primary objective of this presentation is to describe our current 
understanding of the observational properties of black hole X-ray spectra.  
In addition to describing the phenomenology of black hole spectral states, 
I discuss physical implications of the observations, including how they 
may constrain accretion geometries and emission mechanisms.  The flexible 
scheduling and broad bandpass of {\em RXTE} have been critical to these 
studies and especially to our group (Tomsick, Kalemci, Corbel, and Kaaret) 
as we have focused on frequent {\em RXTE} and radio observations close to 
state transitions.  Much of the recent debate on black holes has been 
related to the physics of the Hard State, and, here, I provide a summary 
of how the predictions of the current theoretical models compare to 
observations.  This presentation includes some discussion of black hole 
X-ray timing and radio emission, but these topics are dealt with in more 
detail by others in these proceedings (see papers by Remillard and Corbel). 
Finally, I discuss the instrumental capabilities that will be important 
in future studies of black hole spectra.

\section{Black Hole Systems Observed by {\em RXTE}}

\begin{figure}
  \includegraphics[height=.45\textheight]{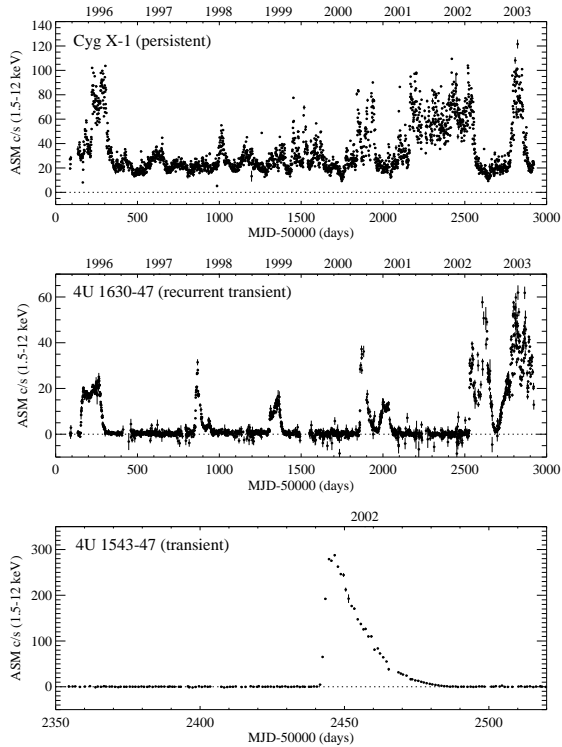}
  \caption{X-ray light curves for accreting black holes from the {\em RXTE} 
All-Sky Monitor.  The energy band is 1.5-12 keV, and each point represents 
the average count rate over 1 day.  The top two panels show the light 
curves for a persistent source (Cyg X-1) and a recurrent transient 
(4U 1630--47) over the entire {\em RXTE} mission to date.  The bottom 
panel shows a single outburst from a transient source (4U 1543--47).
\label{fig:asm}}
\end{figure}

During the {\em RXTE} lifetime (1996-present), 28 accreting black hole
or black hole candidate systems have been observed in outburst.  Compact 
object mass measurements were obtained, via optical or infrared 
observations, for 11 of these systems, and the masses are higher than 
3\Msun, the theoretical mass upper limit for neutron stars, in all of 
these.  Eight of the 28 sources have been in outburst for the entire 
{\em RXTE} mission, and Figure~\ref{fig:asm} shows an example light 
curve for one of the ``persistent'' sources (Cyg X-1).  Four of the 
sources have had multiple outbursts between 1996 and now, and the
light curve for the most active ``recurrent transient'' (4U 1630--47)
is shown in Figure~\ref{fig:asm}.  The other 16 sources have had only
one outburst observed by {\em RXTE}, but it should be noted that several
of these sources had outbursts before 1996 \citep{csl97} and are know to 
be recurrent.

\section{Spectral States}

\begin{figure}
  \includegraphics[height=.45\textheight]{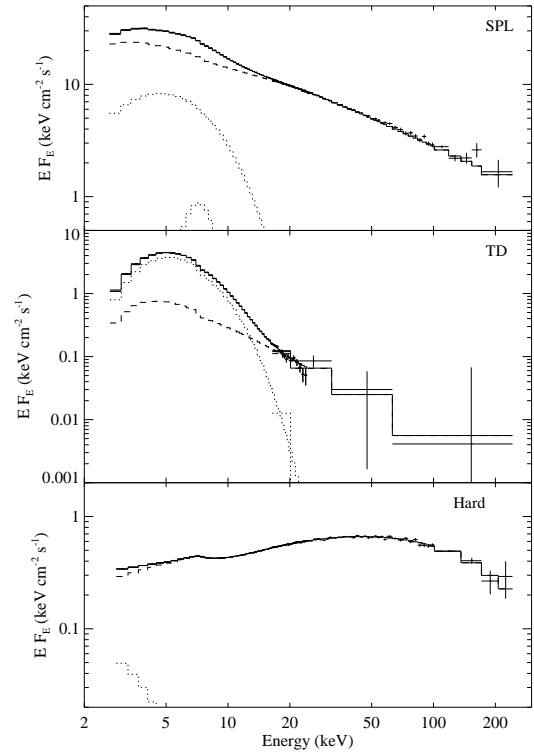}
  \caption{Examples of {\em RXTE} (PCA+HEXTE) energy spectra of black 
hole systems in the Steep Power-Law, Thermal Dominant, and Hard states.  
In each panel, the soft disk component (dotted line), power-law or
cutoff power-law component (dashed line), and the total model spectrum
(solid line) are shown.  The SPL spectrum is from a 1996 observation of
GRO J1655--40.  Note that an iron K$\alpha$ line is also included in 
the model (see the paper by Rothschild et al. in these proceedings).  
The TD spectrum is from a 1996 observation of 4U 1630--47.  The Hard 
spectrum is from a 2001 observation of XTE J1650--500.\label{fig:ss}}
\end{figure}

Although the energy spectra of accreting black holes are complex, they 
are dominated by two emission components: a soft thermal component and 
a hard power-law or cutoff power-law component extending to hundreds of 
keV.  The soft component is very likely blackbody emission from an
accretion disk, and the disk may have properties similar to (or the
same as) a \citet{ss73} disk.  Spectral states 
are defined, in large part, based on the relative strength of the two 
components.  Recently, the large number of {\em RXTE} observations of 
accreting black holes in various spectral states were used to quantitatively 
define the spectral states \citep{mr03}.  In addition to giving the most 
precise spectral state definitions to date, \citet{mr03} considered the 
fact that there are many examples (see below) for which the traditional 
luminosity-based state names are not appropriate and re-named the spectral 
states.  A listing of the three outburst states and their general 
properties follow \citep[see][for the complete quantitative 
definitions of the states]{mr03}.  Figure~\ref{fig:ss} shows examples 
of spectra for sources in these three states.\\
$\bullet$~The {\bf Steep Power-Law (SPL)} state was formerly known as the
Very High state.  Both disk and power-law components are usually present 
in this state, and a source is said to be in the SPL state if either
the power-law accounts for more than 50\% of the 2-20 keV flux or if 
quasi-periodic oscillations (QPOs) are present and the power-law accounts 
for more than 20\% of the 2-20 keV flux.  In this state, the power-law
component has a photon index of $\Gamma > 2.4$.\\
$\bullet$~The {\bf Thermal-Dominant (TD)} state was formerly known as the
High-Soft state.  The disk component dominates the X-ray spectrum, 
accounting for more than 75\% of the 2-20 keV flux.  The level of timing
noise in this state is usually very low.\\
$\bullet$~The {\bf Hard} state was formerly known as Low-Hard 
state\footnote{Even McClintock \& Remillard (2003) continued to use 
the old terminology for this state, but several examples indicate that
the ``Low-'' should be dropped.}.  The power-law, with 
$1.5 < \Gamma < 2.1$, dominates in this state, accounting for more 
than 80\% of the 2-20 keV flux.  There is a high level of timing noise, 
and radio emission, likely from a compact jet, is typical of this state
\citep{fender01}.

An ``Intermediate State'' has also been defined as one of the canonical
black hole spectral states \citep{mendez98}; however, the only property that
clearly separates this state from the SPL state is that the Intermediate 
State has traditionally been used to describe black holes at lower 
luminosity.  Thus, the argument that the states are not strictly
dependent on luminosity makes this state, as canonically defined, 
unnecessary.  However, there are times, especially during state 
transitions, where the source properties do not fit into the SPL, 
TD, or Hard states, and I agree with the suggestion of \citet{mr03}
that the states that show various combinations of SPL, TD, and 
Hard state properties be referred to as intermediate states.  

There is a final state that is simply defined by (low) luminosity 
called the ``quiescent'' state.  For black hole systems, a source is 
typically said to be in quiescence if its X-ray luminosity is in the 
$10^{30-33}$ erg~s$^{-1}$ range \citep{garcia01,hameury03,tomsick03a}.
However, due to the relatively poor statistical quality of the 
low luminosity observations, it is difficult to determine if there 
are other source properties that distinguish quiescent properties 
from Hard state properties, and, below, I discuss this issue further.

\section{Examples of Luminosity Independence of States}

\begin{figure}
  \includegraphics[height=.3\textheight]{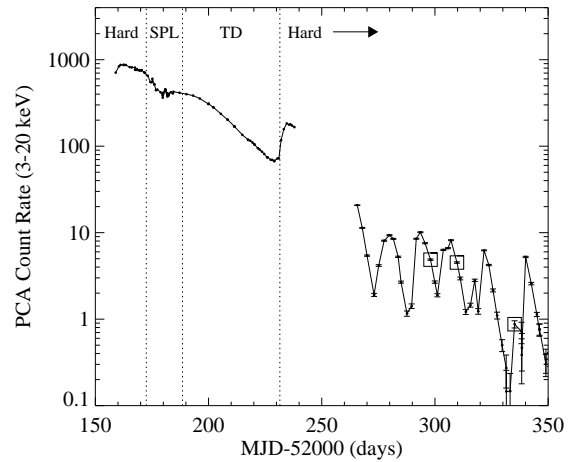}
  \caption{The 3-20 keV light curve for XTE J1650--500 during its
2001-2002 outburst.  Each point shows the Proportional Counter
Array (PCA) count rate from a pointed {\em RXTE} observation.
The gap in coverage centered at MJD 52,250 is due to a sun
angle constraint.  More details about the 14 day oscillations
that occurred after the sun gap are presented in \citet{tomsick03b}.
The three squares during this part of the outburst mark the times
of {\em Chandra} observations.  XTE J1650--500 exhibited Hard state
properties at luminosities ranging over three orders of magnitude.
\label{fig:1650states}}
\end{figure}

Here, I provide some examples of sources that exhibit X-ray properties 
that fit in well with the standard spectral state definitions outlined
above, but do not show the luminosity dependence that is considered to 
be typical.  This is not meant to be an exhaustive list, but I simply 
discuss some of the clearest examples.  It has long been known that
some sources exhibit ``Hard state outbursts'' where they reach high
luminosities, but remain in the Hard state.  The most dramatic example
of this behavior is seen from V404 Cyg (= GS 2023+338), which remained
in the Hard state while reaching a luminosity of 
$\sim$$2\times 10^{39}$ erg~s$^{-1}$, assuming a distance of 3.5 kpc
\citep{ts96}.  A more recent example is the behavior of XTE J1650--500
as this source began its outburst in the Hard state and remained in
the hard state for two weeks (see Figure~\ref{fig:1650states}), during
which time the 3-20 keV luminosity peaked.  The source eventually made
a transition to the SPL and TD states \citep{homan03}, and, later in 
the outburst, the source showed a transition back to the Hard state
\citep{kalemci03,homan03}.  {\em Chandra} observations occurred near 
the end of the outburst, showing that the source exhibited Hard state 
properties at a luminosity of $9\times 10^{34}$ erg~s$^{-1}$ \citep{tkk03}.  
Overall, XTE J1650--500 exhibited Hard state properties over a range
of three orders of magnitude in X-ray luminosity.  Another excellent
example of luminosity independence comes from a study of a full 
outburst from XTE J1550--564 \citep{homan00}.  The hardness-intensity
diagram for XTE J1550--564 indicates that this source can enter the
Hard state, the SPL state, or various intermediate states over a 
large intensity range.  Other examples include GRS 1758--258 and
1E 1740.7--2942, which show transitions from Hard to TD states as
the source luminosity drops \citep{smith01,smith02}.  Also, Cygnus X-1
\cite{zhang97} and SAX J1711.6--3808 \citep{wm02} illustrate the
luminosity independence of spectral states.

In addition to being important for determining the most meaningful
classification of spectral states, this luminosity independence
implies that the states are not simply set by the mass accretion
rate.  As discussed by \citet{homan00} and others, this 
indicates that another physical parameter that is not directly
tied to the mass accretion rate must be important in setting the
emission properties.  Although there have been suggestions for
what this second parameter might be, such as the inner radius
of the optically thick accretion disk ($R_{\rm in}$), the size of 
the corona, or the jet power, it is currently unclear which of 
these parameters cause significant changes in the X-ray emission
properties.

\section{X-Ray Properties and Physical Implications}

In this section, I describe the specific observational properties 
of different phases of the evolution of accreting black holes and 
discuss the physical implications of the observations.  The 
different evolutionary phases include: states that typically
exhibit a significant or dominant soft component (SPL and TD); 
transitions to the Hard state; and the Hard state.

\subsection{Steep Power-Law and Thermal-Dominant States}

The SPL and TD state spectra both typically include a significant
contribution from an optically thick disk, but, with the stronger 
power-law component in the SPL, a third ``reflection'' component 
arises due to the hard flux illuminating the accretion disk 
\citep{lw88}.  The two main features of the reflection component 
are an iron K$\alpha$ emission line near 6.4 keV and a ``reflection 
bump'' between 10 and 100 keV, and, for many broadband black hole
spectra, these features are present \citep[e.g.,][]{gierlinski99}.
Recent observations by {\em Chandra} and {\em XMM-Newton} 
\citep[e.g.,][]{miller02} indicate that many black hole systems 
have an intrinsically broad iron line, and the broadening is 
interpreted as being caused by a redshift due to the black hole's 
gravitational field.  The inner radii inferred from spectral 
modeling of the broad lines provides sone of the strongest 
evidence that the inner radius of the disk reaches the Innermost 
Stable Circular Orbit (ISCO).  Other evidence that the disk extends 
close to the black hole comes from the detection of high frequency 
quasi-periodic oscillations (QPOs), which are only seen in the SPL 
state (see the paper by Remillard in these proceedings), and also 
from modeling the soft spectral component using a \citet{ss73} model 
(although it has been shown that, in some cases, the inner radii 
inferred from such fits give unphysically small values).

While there is a relatively good understanding of the origin 
of the soft component and of the geometry of the optically thick 
portion of the accretion flow, the origin of the power-law 
component is currently unclear.  However, the fact that 
the power-law in the SPL state can extend up to energies
approaching at least 1 MeV \citep[e.g.,][]{tomsick99} indicates 
that a non-thermal electron energy distribution is required.  
Understanding the production of this component in detail is an 
important, unsolved problem (see the paper by Coppi in these 
proceedings for more on this issue).

\subsection{Transitions to the Hard State}

A main focus of our group has been to measure the properties of 
accreting black holes in X-rays with {\em RXTE} and in the radio
band close to state transitions.  Frequent (e.g., daily) monitoring 
is critical for this study because it is known that state 
transitions can occur on time scales of days \citep{tk00}.  The most 
accessible and reliable state transition is the transition from the 
TD or SPL state to the Hard state at the ends of transient outbursts, 
and performing radio observations close to such a transition is
especially interesting for understanding the compact jet turn-on.

Using our observations as well as observations from the {\em RXTE}
archive, \citet{kalemci_thesis} analyzed data from observations of 
16 outbursts from 11 different sources, and, in eight cases, good 
{\em RXTE} coverage was obtained close to the state transition.  
The data are shown in the paper by Kalemci et al. in these proceedings, 
and I summarize some of the basic results here.  The sharpest changes 
were seen in the timing properties, with sources typically showing a 
change from a very low, undetectable, level of timing noise to an RMS 
level of tens of \% in less than a day.  It is clear that the timing 
noise can turn on very rapidly in these systems.  While the timing 
changes are very sharp, gradual changes are seen in the spectral 
parameters, including the power-law photon index, $\Gamma$, and the 
temperature at the inner edge of the accretion disk, $kT_{\rm in}$.  
Finally, a very large drop in the 3-25 keV flux of the soft component 
is apparent, and, typically, this component becomes undetectable after 
the transition.  

\begin{figure}
  \includegraphics[height=.44\textheight]{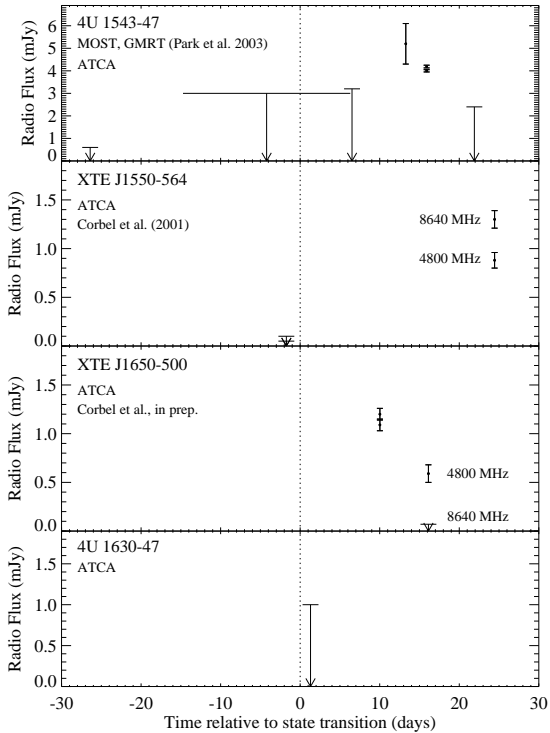}
  \caption{Radio observations of black hole transients made close 
to the times of transitions to the Hard state during outburst decay.
For each source, the dotted line marks the sharp change in the
level of the RMS timing noise (see the paper by Kalemci in these
proceedings).  4U 1543--47 has the best coverage when our radio 
data is combined with data from \citet{park03}.  The radio flux
does not increase until 8-12 days after the increase in the RMS
noise.  The XTE J1550--564 points were previously published in
\citet{corbel01}.\label{fig:radio}}
\end{figure}

The sharp timing transition that is seen provides a clearly defined 
transition time, and Figure~\ref{fig:radio} shows radio measurements 
that were made close to this time for four systems.  For 4U 1543--47, 
XTE J1550--564, and XTE J1650--500, we obtained radio detections after 
the state transition.  For 4U 1630--47, we obtained an upper limit
very close to the time of the state transition.  The most interesting 
case is 4U 1543--47 because, when our data are combined with the
measurements reported by \citet{park03}, very good radio coverage was 
obtained.  For this source, the radio emission increased, but not until 
8-12 days after the state transition (see also the paper by Buxton et al. 
in these proceedings for information on the behavior of 4U 1543-47 
during this time in the infrared).  Our analysis of the {\em RXTE} 
data for 4U 1543--47 indicates that the first radio detection occurred 
about the time that the source reached the true Hard state as indicated 
by $\Gamma$ reaching its hardest value of about 1.6 (Kalemci et al., in 
preparation).  This suggests that the processes that lead to an increase 
in timing noise do not automatically result in jet production, and 
re-enforces the connection between jets and the hardness of the X-ray 
spectrum \citep[e.g.,][]{corbel00}.

\subsection{Hard State}

A current topic of debate concerns the questions of what is 
the X-ray emission mechanism and what is the source geometry
in the Hard state.  The standard picture is that the cutoff 
power-law is produced by inverse Comptonization from a thermal
distribution of coronal electrons where the coronal temperature
is close to the cutoff energy.  While the energy spectrum is
well-described by thermal Comptonization, the discovery that 
a compact jet is present in the Hard state has led to the 
question of whether at least some of the X-ray emission could 
be produced in the jet, and jet-based synchrotron \citep{mff01}
and Compton \citep{gak02} models have been investigated.

One specific issue concerning the source geometry is whether
the inner radius of the optically thick accretion disk 
increases in the Hard state.  One might think that the drop in 
flux and temperature of the soft component during and after the 
transition to the Hard state described above demonstrates that
$R_{\rm in}$ increases; however, other possibilities have been
suggested such as the idea that the inner disk remains intact, 
but, in the Hard state, the viscosity mechanism in the inner
disk turns off and the accretion energy goes into producing
the jet or heating the corona rather than being radiated 
away in the inner disk \citep{lpk03}.  Still, even if the
inner disk does not radiate efficiently, one might expect
to see a reflection component if the inner disk remains at
the ISCO.  Studies of the strength of the reflection component
have been carried out, showing a large decrease in the
strength of the reflection component as $\Gamma$ becomes
harder when the source passes through intermediate states
on its way to the hard state \citep{zls99,zdziarski03}.
While the trend in the strength of the reflection component
is consistent with an increase in $R_{\rm in}$, other 
possible explanations for the trend include ionization of
the inner part of the disk \citep{nayakshin00}, but also see
\citep{bdn03}, or beaming of emission away from the disk 
\citep{beloborodov99}.

\begin{figure}
  \includegraphics[height=.28\textheight]{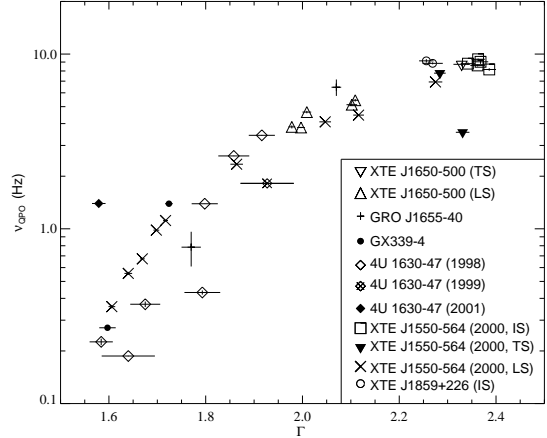}
  \caption{The QPO frequency vs. the power-law photon index
($\Gamma$) for six different black hole transients during outburst 
decay \citep[from][]{kalemci_thesis}.  This illustrates the gradual 
drop in the QPO frequency as the sources enter the Hard state, 
and shows that these sources all lie on approximately the same 
correlation.\label{fig:gamma_frequency}}
\end{figure}

While the evolution of the soft component and the reflection
component do not provide a definitive answer to the question
of how $R_{\rm in}$ evolves, it is notable that changes in
both of these components are well-explained by an increasing
$R_{\rm in}$.  Furthermore, an increasing $R_{\rm in}$ can
provide a rather natural explanation for observed changes
in timing properties.  As shown in Figure~\ref{fig:gamma_frequency}, 
the black hole sources appear to show a uniform drop in
QPO frequency as $\Gamma$ hardens \citep{kalemci_thesis}, and 
similar trends have been reported for either QPO frequencies
or break frequencies in the power spectrum in other work
\citep{dp99,nwd02,rgc01,vignarca03}.  An increasing
$R_{\rm in}$ may help to explain these trends as it provides
an increasing dynamical time scale in the system.  

\begin{figure}
  \includegraphics[height=.28\textheight]{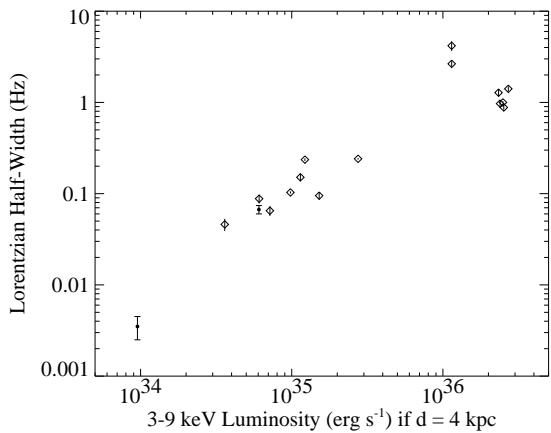}
  \caption{The Lorentzian Half-Width (approximately the
``break frequency'' described in the text) for the
continuum component in the power spectrum vs. the source
luminosity for the black hole candidate XTE J1650--500
in the Hard state \citep[see][]{tkk03}.  Observations
come from {\em RXTE} (diamonds) and {\em Chandra}.
\label{fig:ff_1650}}
\end{figure}

Recently, we used {\em RXTE} and {\em Chandra} observations of
XTE J1650--500 to show that the trend of characteristic
frequencies dropping with luminosity extends to very low
luminosities near $1\times 10^{34}$ erg~s$^{-1}$ as shown in
Figure~\ref{fig:ff_1650} \citep{tkk03}.  We detect a
change in the break frequency by a factor of $\sim$1200
between the transition to the Hard state and the lowest
luminosity we sampled, and if this is set by a dynamical
time scale that is determined by the inner radius of the
disk, the implication is that $R_{\rm in}$ increases by 
a factor of $\sim$110.  The power spectra we measure
for XTE J1650--500 at low luminosities are consistent with 
the standard broken power-law or zero-centered Lorentzian 
models that are typically used to fit Hard state power
spectra \citep{tkk03}, and it is interesting that optical 
observations of other black hole systems in quiescence show 
similar power spectra but with even lower break frequencies 
\citep[see][and the paper by Shahbaz et al. in these 
proceedings]{hynes03}.  Although we are comparing optical 
and X-ray properties, the similar timing properties suggest 
that black hole transients have the same overall structure 
and that the same physical processes are operating in the 
Hard state and in quiescence \citep{tkk03}.  Furthermore, 
the fact that the quiescent energy spectra of black holes
exhibit a component at optical and UV energies that may be 
caused by thermal emission from a truncated disk
\citep{mcclintock03} underscores the idea that physical
properties change gradually as the source luminosity
drops in the Hard state and quiescence but that the
overall structure does not undergo a major transition.

\section{Theoretical Models for the Hard State}

Table~\ref{tab:models} compares the predictions of three different
theoretical pictures for the Hard State to observational properties.  
In the ``Sphere+Disk'' model, the accretion disk is truncated at some 
value of $R_{\rm in}$ and a spherical corona is present for radii 
closer to the black hole \citep{dove97,emn97}.  The ``Slab'' model is 
typically envisioned as an optically thick disk extending close to the 
ISCO with active regions above the disk acting as the corona to produce 
the hard X-ray emission \citep{grv79}.  While the emission mechanism 
for these first two models is inverse Componization, the third, which 
is the X-ray jet model, produces emission via a synchrotron mechanism 
\citep{mff01}.

With the additional free parameter, $R_{\rm in}$, the Sphere+Disk
(SD) model provides a natural explanation for the reflection strength 
vs. $\Gamma$ and characteristic frequency vs. $\Gamma$ correlations 
described above.  In addition, the luminosity independence of states 
argues for the addition of a parameter that may not be directly tied 
to mass accretion rate.  However, while there is observational evidence 
that jets are present in the Hard state\footnote{Steady, compact radio 
jets have been resolved in the Hard state for GRS 1915+105 and Cyg X-1
\citep{dhawan00,stirling01}.  The radio emission from other Hard state 
systems is likely also from unresolved jets.}, jets have not been 
incorporated into SD models such as the Advection-Dominated Accretion 
Flow (ADAF) model.  It has previously been suggested that accretion 
energy in the Hard state might be used to power a jet \citep{bb99,fgj03}, 
but it is not clear if, e.g., the Advection-Dominated Inflow-Outflow 
Solution (ADIOS) can explain the Hard state radio emission.

For the Slab model, it has been demonstrated in more detail that a 
magnetic corona can drive a jet \citep{mf02}.  However, in this model, 
explanations for the reflection and frequency vs. $\Gamma$ correlations 
are unclear.  Perhaps the size of the corona changes, but more work is 
necessary to determine if the correlations can be reproduced within this 
model.  A significant point in favor of the X-ray jet model is that it 
can explain, in detail, the reported correlations between X-ray and 
radio flux \citep[see][and the paper by Corbel in these 
proceedings]{gfp03}; however, similar X-ray/radio flux correlations
may arise in more general situations with non-X-ray-emitting 
jets \citep{hs03}.  More work is necessary to determine if
a synchrotron X-ray jet is viable.  As shown in \citet{zdziarski03}, 
a synchrotron cutoff is expected to be more gradual than the
exponential cutoffs that are typically observed in black hole
spectra.  Also, as discussed in \citet{tkk03}, it is not
clear if the X-ray jet model can reproduce a change in the
characteristic frequencies of the system by a factor of 1200
since this would seem to imply a very large change in the location 
of the shock that produces the high energy electrons that give
rise to X-ray emission.

\begin{table}
\caption{Comparison between Models and Observational Properties\label{tab:models}}
\begin{tabular}{cccccc} \hline
      & Spectral          & Spectral          & Correlations       &     &   \\
      & Continuum        & Continuum        & (Reflection and    &     & Radio/X-Ray \\
Model & (2-20 keV)      & (20-200 keV)    & Frequency)         & Jet & Correlation \\ \hline
Sphere+Disk (e.g., ADAF) & & & & & \\
\citet{dove97} & & & & & \\
\citet{emn97} & Yes & Yes & Yes & No (ADIOS?) & ?\\ \hline
Slab (e.g., Magnetic Corona) & & & & & \\
\citet{grv79} & & & & & \\
\citet{mf02} & Yes & Yes & ? & Yes & Yes?\\ \hline
X-Ray Emitting Jet & & & & & \\
\citet{mff01} & Yes & ? & No? & Yes & Yes\\ \hline
\end{tabular}
\end{table}

\section{Beyond {\em RXTE}}

Studies of black hole X-ray spectra require a broad bandpass
as well as flexible scheduling due to the rapid evolution that 
these sources can exhibit.  With the combination of an all-sky 
monitor, instrumentation (PCA and HEXTE) covering the 3-200 keV 
energy range, and a dedicated Project Scientist and scheduling
team, {\em RXTE} has been an excellent tool for research on
black hole spectra.

An instrument with an even larger effective area, as has been
discussed at this meeting, would be useful for spectral studies.
Black hole systems are highly variable, and a larger area would 
allow for high quality spectra to be obtained over a shorter 
period of time.  For example, it may be possible to study spectral 
evolution on sub-second time scales when QPOs are present.  Also, 
a larger effective area above $\sim$50 keV would allow for detailed 
studies of the evolution of the high energy component.  Although I 
have not discussed measurements of time lags and coherence, our 
group's observations of sources during state transitions indicate 
that these parameters provide useful information on the physics of 
the system \citep{kalemci03}, and a larger area would be very 
beneficial to such measurements.

The spectra shown in Figure~\ref{fig:ss} show that the {\em RXTE}
bandpass captures the two primary spectral components, but it 
should be noted that important information would be lost with a
reduced bandpass.  Background is another consideration as the
transition to the Hard state during outburst decay can occur at
3-25 keV flux levels below $1\times 10^{-9}$ erg~cm$^{-2}$~s$^{-1}$
even for relatively nearby systems with distances of a few kpc.
A significantly higher background level could limit black hole
studies to the brightest portions of the outburst.  Finally, as
most X-ray binaries are in the Galactic plane, source confusion
is a concern for studying fainter sources, and this can be 
problematic for {\em RXTE}, given the $1^{\circ}$ radius
(FWZI) field of view.  A smaller field of view could be beneficial 
for studies of fainter sources in the Galactic plane.

%%%%%%%%%%%%%%%%%%%%%%%%%%%%%%%%%%%%%%%%%%%%%%%%
%% BACKMATTER
%%%%%%%%%%%%%%%%%%%%%%%%%%%%%%%%%%%%%%%%%%%%%%%%

\begin{theacknowledgments}

I would like to thank my {\em RXTE} collaborators Emrah Kalemci, 
Stephane Corbel, and Philip Kaaret, and I am especially grateful
to Jean Swank for all of her hard work and useful suggestions 
while planning and co-ordinating our rather demanding observing 
program.  I acknowledge partial support from NASA grant NAG5-13055.

\end{theacknowledgments}

%% If the bibliography is produced without BibTeX comment out the


%\bibliography{refs}

\end{document}